\documentclass[12pt,a4paper]{article}
\usepackage{jheppub}  
\usepackage{amssymb} 
\usepackage{amsmath}
\usepackage{mathtools}
\usepackage{amsfonts}    
\usepackage{dsfont}
\usepackage{pdfpages}
\usepackage{verbatim}
\usepackage{tensor}
\usepackage{mathrsfs}
\usepackage{textgreek} 
\usepackage[mathscr]{euscript}
\usepackage[normalem]{ulem}
\usepackage{tikz}
\usetikzlibrary{decorations.pathreplacing, decorations.markings,calc,shapes.misc,decorations.pathmorphing,patterns.meta, math}

\newcommand{\ba}{\begin{align}}

\newcommand{\be}{\begin{equation}}
\newcommand{\ee}{\end{equation}}
\def\bd{\begin{tikzpicture}}
\def\ed{\end{tikzpicture}}

\allowdisplaybreaks[1]

\newcommand{\beq}{\begin{equation}}
\newcommand{\eeq}{\end{equation}}
 
\newcommand{\bea}{\begin{eqnarray}}
\newcommand{\ea}{\end{eqnarray}}
\newcommand{\barr}{\begin{array}}
\newcommand{\earr}{\end{array}}

\def\d{{\partial}}

\renewcommand\d{\text{d}}

\title{New insights on near-extremal black holes}

\author{Gustavo J.\ Turiaci${}^{1,2}$}
\affiliation{${}^1$Institute for Advanced Study, Princeton, NJ, USA\\
${}^2$Physics Department, University of Washington, Seattle, WA, USA}

\abstract{We describe two puzzles that arise from a semiclassical treatment of near-extremal black hole thermodynamics. Both puzzles are resolved by realizing that quantum corrections become arbitrarily large at low temperatures, and we explain how the spectrum and dynamics of near-extremal black holes are modified. This analysis also implies that without low energy supersymmetry, such as in the real world, extremal black holes at exactly zero temperature do not exist since the classical picture breaks down completely. In the context of supergravity the analysis is modified; supersymmetric extremal black holes do exist and they are separated from the non-extremal spectrum by a gap power-law suppressed in the entropy. This justifies black hole microstate counting performed in the 90's using string theory. \\ 
~\\
~\\
~\\
\emph{Article written for the ICTS Newsletter.}}

\begin{document}

\maketitle

\section{Introduction: Black holes and quantum gravity}

General relativity and quantum mechanics are the most successful theories describing the real world, each verified in very different regimes. Put together, the two theories seem incompatible. Two physical phenomena arise for which reconciling these theories is crucial. The first is the big bang. The second concerns black holes, the topic of this article. 

A lesson supported by string theory, the leading candidate of a theory of quantum gravity, is that a black hole behaves as a quantum system from the point of view of an observer that remains outside of it. This conjecture is behind the developments of holography and AdS/CFT dualities \cite{Maldacena:1997re,Witten:1998qj,Gubser:1998bc}, which have been extensively tested in the past decades. Assuming that black holes in our world are described by quantum systems, it is indispensable to investigate the rules of quantizing gravity necessary to exhibit such behavior.

A first observation is the success of the `gravitational path integral' pioneered by Gibbons and Hawking \cite{Gibbons:1976ue}. According to this proposal, we first analyze the region exterior to the black hole where gravity is weak and decide which observable we want to study. An example is the black hole thermal partition function or the time dependence of correlation functions between probes sent to the black hole. This choice determines a boundary condition far from the black hole, and one then integrates over all smooth spacetimes and matter configurations near the black hole consistent with the given boundary conditions. 

In quantum mechanics the path integral is equivalent to the Hilbert space approach. In the presence of gravity this is not so evident: there are multiple situations where the gravitational path integral is in apparent tension with the interpretation of the black hole as a quantum system with discrete microstates. Upon closer inspection, most of these discrepancies are removed by a more complete evaluation of the gravitational path integral. In this article we will explain one example concerning near-extremal black holes, understood thanks to developments in Jackiw-Teitelboim (JT) gravity \cite{Jackiw:1984je,teitelboim1983gravitation,Mertens:2022irh} which is amenable to quantization.

\section{Extremal black holes}

In asymptotically flat four dimensional spacetime, black hole geometries are described by only a few measurable parameters: the mass $M$, the angular momentum $\vec{J}$ and the charge $Q$\footnote{We measure charge in units where the Coulomb force between two charges at a distance $d$ has magnitude $F=Q^2/d^2$.}
. This is known as the `no-hair theorem' of black holes. These parameters are not all independent from each other. For a given value of $J=|\vec{J}|$ and $Q$, there is a minimal possible mass, the `extremal mass' $M_{\rm ext}(Q,J)$ such that\footnote{The extremal mass for arbitrary charges is more complicated and not particularly enlightening $M_{\rm ext}(J,Q) = \frac{\sqrt{2} c J}{\sqrt{G_N}\sqrt{ \sqrt{Q^4 + 4 c^2 J^2}- Q^2}}$.}
\beq
M\geq M_{\rm ext}(J,Q) = \begin{cases}
			\frac{|Q|}{\sqrt{G_N}}, & \text{for $J=0$.}\\
			~\\
            \sqrt{ \frac{ c J}{G_N} }, & \text{for $Q=0$.}
		 \end{cases}
\eeq
A black hole saturating this bound is called \emph{extremal}, and black holes that are close to saturating it are \emph{near-extremal}. As we tune the mass of the black hole below $M_{\rm ext}$, the event horizon disappears leaving behind a naked singularity. This could hardly be called a black hole, and the singularity would represent a lack of predictability of the theory, thus ruling out all solutions with $M<M_{\rm ext}$ as unphysical. The conjecture that this interpretation is correct and all singularities are protected by event horizon is the `cosmic censorship conjecture.'

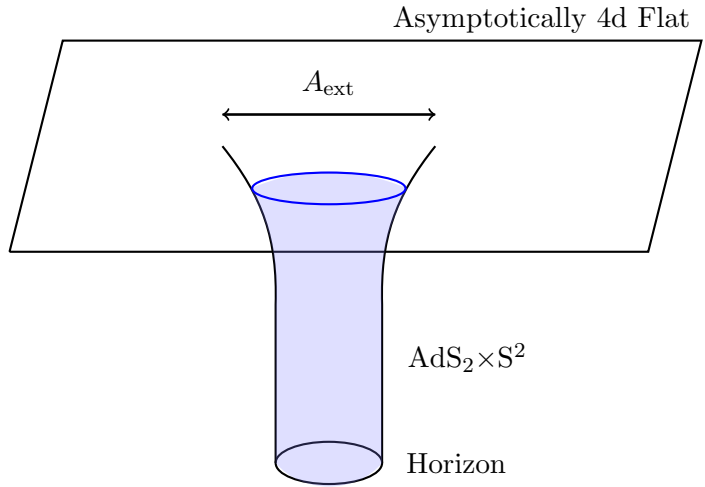
\begin{figure*}
     \centering
   \begin{tikzpicture}[scale=1.4, baseline={([yshift=0cm]current bounding box.center)}]
\draw[thick] (0,-2) ellipse (0.5 and 0.2);
\draw[thick] (-1,1) to [bend left=20] (-0.5,-0.5) -- (-0.5,-2) ;
\draw[thick] (1,1) to [bend right=20] (0.5,-0.5) -- (0.5,-2) ;
\draw[thick] (-3,0) -- (3,0) -- (3.5,2) -- (3.5-6,2) -- (-3,0);
\draw (2,2.2) node {\small Asymptotically 4d Flat};
\draw (1.2,-2) node {\small Horizon};
\draw[->,thick] (-1,1.3) -- (1,1.3);
\draw[->,thick] (1,1.3) -- (-1,1.3);
\draw (0,1.6) node {\small $A_{\rm ext}$};
\draw (1.3,-1) node {\small AdS${}_2\times$S${}^2$};
\fill[blue!50,nearly transparent] (-.74,.6) to [bend left=15] (-0.5,-0.5) -- (-0.5,-2) to [bend right=50] (.5,-2) -- (.5,-.5) to [bend left=15] (.74,.6) to [bend right =22] (-.74,.6);
\draw[thick,blue] (0,.6) ellipse (0.72 and 0.15);
\end{tikzpicture}
     \caption{\footnotesize Spatial geometry of a near-extremal black hole. The throat is shown in blue. At the bottom of the $AdS_2 \times S^2$ throat (shaded region) is the event horizon. The dynamics in this region is described by Jackiw-Teitelboim gravity coupled to matter. Exterior to the blue line is the asymptotically flat spacetime. }
     \label{fig:ThroatDiagram}
 \end{figure*}

Black holes have few isometries (meaning transformations that leave the geometry invariant). For generic values of $(M,Q,\vec{J})$ they correspond to time translations and rotations around the $\vec{J}$ axis. The first reason that near-extremal black holes are interesting is that a powerful new symmetry emerges near the horizon: scale invariance. Extremal black holes develop a long throat near the horizon corresponding to two dimensional Anti-de Sitter space along the time and radial directions, $AdS_2$, fibered over the angular directions, see fig. \ref{fig:ThroatDiagram}. $AdS_2$ is a special space since it is invariant under simultaneous rescalings of the time and radial coordinate. This isometry is actually enhanced to the full conformal group called ${\rm SL}(2,\mathbb{R})$. Several aspects of the dynamics of near-extremal black holes are controlled by this emergent symmetry, which is softly broken close to extremality.
 
 The second reason near-extremal black holes are special is the following. To make the conjecture that black holes are quantum systems precise, we need to separate a spacetime region that we identify with the black hole quantum system from the environment. This separation becomes sharpest near extremality -- the black hole quantum system describes the $AdS_2$ throat.  

Another reason concerns Hawking radiation \cite{Hawking:1975vcx}. Black holes are thermal objects and radiate at a temperature $T$ that depends on $M$, $Q$ and $J$ in a known way. In the near-extremal limit the temperature is low and vanishes at extremality. We define near-extremal black holes more precisely when the temperature is smaller than $k_B T \ll \hbar c/ \sqrt{A}$, where $A$ is the area of the event horizon. Therefore while generic black holes evaporate, near-extremal ones do so very slowly. (Of course, other effects might also lead to instabilities such as superradiance or Schwinger pair production near the horizon that produce a discharge of the black hole.) These features make an understanding of black hole microstates in this regime more likely. 

Finally, studying the near-extremal limit of rapidly rotating black holes that are uncharged $Q=0$ could potentially be relevant for astrophysical black holes.

\section{Two puzzles}

Near-extremal black holes are quite subtle. We shall describe two puzzles about them which were raised long time ago, but only recently addressed.

\begin{enumerate}
\item Extremal black holes have the minimal possible mass given $(Q,J)$, and have zero temperature. They would therefore correspond to the ground state(s) of the putative quantum system describing them. Another property characterizing them is their large entropy! The Bekenstein-Hawking entropy, as derived from a semiclassical analysis of the gravitational path integral, is proportional to the area $A$ of the event horizon measured in units of the Planck length $ \ell_{\rm Pl} = \sqrt{G_N \hbar/c^3}$. This is a very large quantity for macroscopic objects such as a black hole. At extremality even though $T=0$ the area of the event horizon remains large
\beq
S_{\rm ext}= k_B\, \frac{A_{\rm ext}}{4 \ell_{\rm Pl}^2} = k_B\,\frac{\pi \sqrt{Q^4 + 4 c^2 J^2}}{c \hbar}.
\eeq  
For illustration, an extremal black hole with the same spin-to-mass-ratio as M87 has $S_{\rm ext}/k_B \sim 10^{70}$. Such a large zero-temperature entropy violates Nernst's third law of thermodynamics. In one of its formulations, this law claims that the entropy of a system must vanish in the zero temperature limit. This is not a theorem, but a phenomenological observation: the statistical mechanics interpretation of the entropy at zero temperature is as a ground state degeneracy which in the absence of any symmetry is expected to be small. According to classical black hole thermodynamics, the quantum system that describes a near-extremal black hole has an extensive number of ground states, with respect to the number of degrees of freedom $N\sim O(S_{\rm ext})$, and therefore violates the third law. One therefore is forced to either find an argument that removes this degeneracy, or find a symmetry principle that explains it. This issue was emphasized by Page \cite{Page:2000dk}.

\item The second puzzle was first raised by Preskill, Schwarz, Shapere, Trivedi and Wilczek in 1991 \cite{Preskill:1991tb}, and further elaborated by Maldacena, Michelson and Strominger in 1998 \cite{Maldacena:1998uz}. The thermal treatment of black holes is appropriate if the emission of a typical quantum of radiation does not change the temperature by a substantial amount. Preskill \textit{et al.} realized this property is lost for near-extremal black holes when the temperature becomes low enough. The temperature change upon emission of a Hawking quanta is given by 
\beq
\frac{\delta T}{T} = \frac{k_B}{c^2}\left| \left(\frac{\partial T}{\partial M}\right)_{Q,J} \right|,
\eeq
When the right-hand-side becomes small, the thermal description breaks down. This happens for temperatures lower than\footnote{At this point, the order one coefficient in $T_{\rm breakdown}$ is arbitrary.}
\beq
T_{\rm breakdown} =\frac{\pi c^3 \hbar}{G_N M_{\rm ext}}\,\frac{1}{S_{\rm ext}} = \frac{\pi  \hbar^2}{\ell_{\rm Pl}^2 M_{\rm ext}}\,\frac{1}{S_{\rm ext}} .
\eeq
For macroscopic black holes, this is extremely small $T_{\rm breakdown}\sim O(1/S_{\rm ext})$. For a black hole with the same spin-to-mass-ratio as M87 it is of order $T_{\rm breakdown} \sim 10^{-120} {\rm K}$. Examples from string theory suggested that there is a gap of order $E_{\rm breakdown}=k_B T_{\rm breakdown}$ in the energy spectrum of near-extremal black holes thus removing the problematic states, although there was no calculation in gravity supporting this claim. Moreover, this gap would be unexpectedly large, power-law suppressed in the entropy, while for a chaotic system such as the black hole spectrum, gaps are expected to be exponentially small in $S_{\rm ext}$. What would cause such a large gap?

\end{enumerate}

Both puzzles are resolved by taking the gravitational path integral seriously. When evaluating it near extremality, there are certain gravitational modes that become very light at low temperatures. Their quantum fluctuations therefore cannot be ignored and the classical picture that lead to these two puzzles is strongly modified. This was realized thanks to recent developments in Jackiw-Teitelboim gravity, which we explain next.

\section{Jackiw-Teitelboim gravity: A resolution}

The geometry of near-extremal black holes develops a long throat described by an $AdS_2$ space fibered over the angular coordinates of $S^2$. As an illustration, if one wants to study  scattering of a probe off the black hole, it is natural to treat the throat and the exterior region separately. In the exterior region the probe is far from the black hole and gravity is weak. When the probe reaches the throat, the interaction with the black hole is important and the $AdS_2$ description becomes useful. The dynamics of gravity and matter on $AdS_2 \times S^2$ can be conveniently repackaged as a two dimensional theory on $AdS_2$ as follows (see for example \cite{Gaikwad:2018dfc,Nayak:2018qej,Sachdev:2019bjn})
\begin{itemize}
\item \textbf{JT gravity:} This is a 2d theory of dilaton-gravity that describes the dynamics of spherically symmetric fluctuations of the $AdS_2$ metric, and spherically symmetric fluctuations of the total area of the transverse sphere $S^2$. From the 2d point of view, the latter mode is a scalar field called the `dilaton.' 
\item \textbf{2d Matter:} There are two types of 2d matter fields that arise from the higher-dimensional theory. The first corresponds to spherically symmetric modes of light matter that were already present in four dimensions. The second corresponds to modes with non-trivial angular dependence coming either from higher-dimensional light matter or from the higher-dimensional metric itself. In 2d both sets of fields are described in a unified way.  
\end{itemize}
The matter content that appears in $AdS_2$ can be quite complicated. Since the size of the sphere $S^2$ is of the same order of magnitude as the size of $AdS_2$, modes with non-trivial angular dependence cannot be integrated out since they are not heavy -- we are left in 2d with a large number of light fields. The simplification instead arises because interactions between JT gravity and light matter become very simple\footnote{In more detail, the matter couples minimally to the 2d metric but to leading order does not couple to the dilaton. This simplification is crucial.} and even solvable \cite{Bagrets:2017pwq,Stanford:2017thb, Mertens:2017mtv}! A recent review on these developments can be found in \cite{Mertens:2022irh}.

A non-trivial fact understood only recently (thanks to parallel developments in condensed matter systems such as the Sachdev-Ye-Kitaev models \cite{Chowdhury:2021qpy}) is that JT gravity has two coupling constant that should be considered independent. The first is $G_N$ which is the obvious one -- gravity is weak when Newton's constant $G_N$ is small. The second and more subtle one is the temperature itself \cite{Almheiri:2014cka,Jensen:2016pah,Maldacena:2016upp,Engelsoy:2016xyb}. Quantum effects become large when the temperature is low and small when the temperature is high. The transition temperature derived from the JT gravity description of the higher-dimensional black hole is at precisely the same scale $T_{\rm breakdown}$ identified by Preskill \emph{et al}.

Intuitively, quantum effects captured by JT gravity arise from a mode that become light at extremality: time-dependent fluctuations of the length of the throat. As the temperature is lowered quantum fluctuations are less and less suppressed. Besides characterizing this mode, recent developments in JT gravity explain how to quantize it exactly! This is true even in the presence of matter thanks to the simplifications alluded above. For simplicity, we illustrate this for $J=0$. The quantum-corrected near-extremal entropy (for $k_B T \ll \hbar c/ \sqrt{A_{\rm ext}}$) becomes\footnote{A version of this result was first derived in three dimensions \cite{Ghosh:2019rcj,Maxfield:2020ale}. Afterwards this was extended to four dimensions \cite{Iliesiu:2020qvm, Iliesiu:2022onk} in a way easily generalized to higher dimensions \cite{Boruch:2022tno}. Similar quantum effects arise in the Nariai limit of black holes in de Sitter and were studied in \cite{Maldacena:2019cbz}.}
\bea
\frac{S(T)}{k_B }&\approx& \underbrace{\frac{A_{\rm ext}}{4\ell_{\rm Pl}^2} + \frac{4\pi^2 T}{T_{\rm breakdown}}}_{\text{Classical Bekenstein-Hawking entropy}}\nonumber\\
&&\hspace{-0.5cm}+\underbrace{ \underbrace{\left(\frac{-n_S - 62 n_V - 11 n_F-964}{180}\right)}_{=c_{\rm log}}\log  \left(\frac{A_{\rm ext}}{4\ell_{\rm Pl}^2}\right)+ \underbrace{ \frac{3}{2} \log \left( \frac{T}{T_{\rm breakdown}}\right)}_{\text{JT mode}}}_{\text{Quantum Corrections} } .\label{eq:Sqtmc}
\ea
The first two terms come from classical gravity. The last two arise from quantum corrections to the gravitational path integral. The temperature-independent correction gets contributions from all fields and depends on the number of 4d light scalars $n_S$, vectors $n_V$, and Dirac fermions $n_F$. Its evaluation was pioneered by Sen \cite{Sen:2008vm,Sen:2011ba}. Importantly, the last term is the only temperature-dependent quantum correction and comes from JT gravity alone, making it universal. 

These considerations address the puzzle raised by Preskill \emph{et al.} -- regardless of how small $G_N$ is, when the temperature is low enough the quantum effects from the JT mode will be unavoidably large. When $T\lesssim T_{\rm breakdown}$ the log-T correction in eqn.~\eqref{eq:Sqtmc} dominates over the classical linear-in-$T$ contribution from the first line. Since quantum corrections are large, the classical analysis is no longer applicable. This also addresses the first puzzle. The quantum-corrected entropy becomes small (order one) at an even lower temperature scale $T \sim T_{\rm breakdown} \exp{(-\frac{A_{\rm ext}}{4 \ell_{\rm Pl}^2})}$. At such ultra-low temperatures other non-perturbative corrections to the gravitational path integral can compete with the black hole saddle and the actual ground state can be quite complicated. The important conclusions are that (i) the prediction from gravity is consistent with an order one number of ground states and (ii) the ground state(s) are not in any way described by an extremal black hole since the classical description is completely lost at $T=0$.  
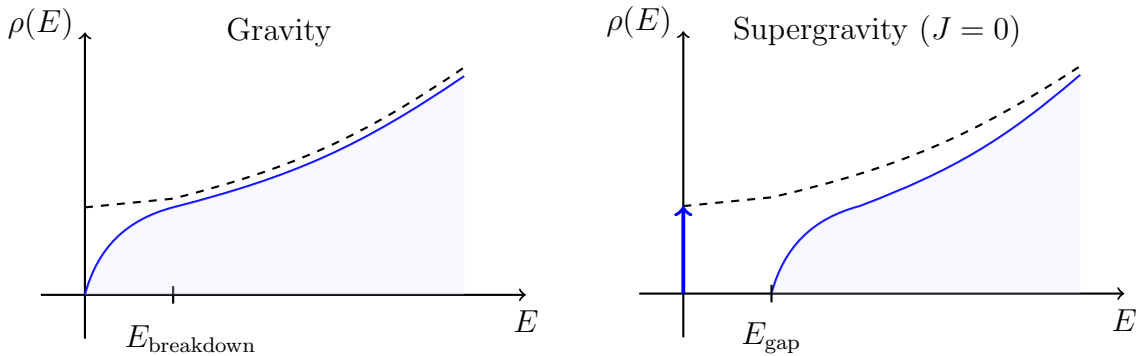
\begin{figure}[t]
    \centering
\begin{tikzpicture}[scale=1.16, baseline={([yshift=0cm]current bounding box.center)}]
\draw[thick,->] (0,-0.5) -- (0,3);
\draw[thick,->] (-0.5,0) -- (5,0);
\draw[thick, blue ] (0,0) to[bend left =30] (1,1) to[bend right = 10] (4.3,2.5) ;
\draw[thick, dashed] (0,1) to[bend left =0] (1,1.1) to[bend right = 10] (4.3,2.6) ;
\node at (5,-0.3) {$E$};
\node at (2.2,3) {Gravity};
\node at (-0.5,3.1) {$\rho(E)$};
\draw[thick] (1,-0.1) -- (1,+0.1);
\node at (1.2,-0.5) {$E_{\rm breakdown}$};
\fill[blue!10,nearly transparent]  (0,0) to[bend left =30] (1,1) to[bend right = 10] (4.3,2.5)-- (4.3,0) -- (0,0);
\end{tikzpicture}
~~~
\begin{tikzpicture}[scale=1.16, baseline={([yshift=0cm]current bounding box.center)}]
\draw[thick,->] (0,-0.5) -- (0,3);
\draw[thick,->] (-0.5,0) -- (5,0);
\draw[thick, blue ] (1,0) to[bend left =30] (2,1) to[bend right = 10] (4.5,2.5) ;
\draw[thick, dashed] (0,1) to[bend left =0] (1,1.1) to[bend right = 10] (4.5,2.6) ;
\node at (5,-0.3) {$E$};
\node at (2.2,3) {Supergravity $(J=0)$};
\node at (-0.5,3.1) {$\rho(E)$};
\draw[thick] (1,-0.1) -- (1,+0.1);
\node at (1,-0.5) {$E_{\rm gap}$};
\fill[blue!10,nearly transparent]  (1,0) to[bend left =30] (2,1) to[bend right = 10] (4.5,2.5)-- (4.5,0) -- (0,0);
\draw[line width = .5mm, blue, ->] (0,0) -- (0,1) ;
\end{tikzpicture}
    \caption{(a) Density of states for a near-extremal black hole with fixed $Q$ and $J$, as a function of $E=Mc^2-M_{\rm ext}c^2$. The dashed line is the classical prediction from gravity. The blue line is the quantum corrected one which strongly deviates from the dashed line as extremality is approached. There is no gap visible in this approximation, and extremal black holes disappear since the density of states vanishes. This spectrum also qualitatively applies to supergravity when $J\neq 0$. (b) In supergravity, if the extremal limit preserve some supersymmetries (which happens when $J=0$) the quantum corrected spectrum displays a gap and the extremal black holes survive with their large classical entropy $S_{\rm ext}$, justifying microstate counting in string theory. }
    \label{fig:my_label}
\end{figure}

In the real world the electron exists with a small enough mass-to-charge-ratio which allows charged extremal black holes to decay. This effect should then be included in the gravitational path integral as well as the quantum effects we focused on. As emphasized by the `weak gravity conjecture' \cite{Arkani-Hamed:2006emk} this implies that there is no truly stable ground state of a charged black hole.

It is instructive to present the density of black hole microstates, shown in fig. \ref{fig:my_label}(a). In terms of the energy above extremality $E = M c^2 - M_{\rm ext} c^2$, the density of states $\rho(E)$, defined through the partition function by $Z(T) = \int \d E \, \rho(E)\, e^{-\frac{E}{k_B T}}$, is given by
\bea
\rho(E)&\approx& e^{\frac{A_{\rm ext}}{4\ell_{\rm Pl}^2} + c_{\rm log} \log  (\frac{A_{\rm ext}}{4\ell_{\rm Pl}^2})} \, (E_{\rm breakdown})^{-1} \sinh \left( \sqrt{  \frac{8\pi^2E}{E_{\rm breakdown}}}\right) \\
&\approx &e^{\frac{A_{\rm ext}}{4\ell_{\rm Pl}^2} + c_{\rm log} \log  (\frac{A_{\rm ext}}{4\ell_{\rm Pl}^2})}(E_{\rm breakdown})^{-1} \times \begin{cases}
			e^{\sqrt{  \frac{8\pi^2 E}{E_{\rm breakdown}}}}, & \text{$E\gg E_{\rm breakdown}$,}\\
			~\\
       \sqrt{  \frac{8\pi^2E}{E_{\rm breakdown}}}, & \text{$E\ll E_{\rm breakdown}$.}
		 \end{cases}
\ea
While for $E\gg E_{\rm breakdown}$ the density of states grows exponentially with energy, consistent with the classical Bekenstein-Hawking entropy, the density of states vanishes at extremality. At energies $E\sim E_{\rm breakdown} \exp{(-\frac{A_{\rm ext}}{4 \ell_{\rm Pl}^2})}$ non-perturbative corrections are large and the semiclassical black hole picture near the horizon is not reliable anymore.

The conclusions are universal and only depend on the pattern of symmetry breaking of a near-extremal black hole. The JT mode is equivalent to the Schwarzian theory, the Goldstone mode that arises from the breaking of conformal invariance by finite temperature effects. We expect this near-extremal spectrum to be valid in full generality, although $A_{\rm ext}$ and $T_{\rm breakdown}$ can depend on the model.  This mode also controls quantum corrections to matter correlators and other dynamical features reviewed in \cite{Mertens:2022irh}.

\subsection{Supersymmetric black holes and string theory}

String theory has provided several examples of specific black holes and their quantum systems, in the context of supergravity. When the extremal black hole preserves supersymmetry we can count microstates and compare with the $S_{\rm ext}$, an approach initiated by Strominger and Vafa in 1996 \cite{Strominger:1996sh} see also the review \cite{David:2002wn}. In asymptotically flat 4d supergravity, this occurs when $J=0$. This raises two questions that were never addressed until now: 
\begin{itemize}
\item Why should we trust the classical formula for the entropy at extremality? 
\item Can we reliably identify a Hilbert space of extremal black holes if gaps between states are not visible semiclassically? 
\end{itemize} 
Again, we resort to the gravitational path integral and its JT gravity formulation. When supersymmetry is present at extremality, new fermionic light modes that modify the quantum corrections to the spectrum appear. JT gravity is generalized to JT supergravity and the result \cite{Heydeman:2020hhw} is shown in fig. \ref{fig:my_label}(b). While the gravity theory is only changed by the inclusion of fermions, the quantum corrected spectrum is now completely different! A gap is generated given, to leading order in small $G_N$ expansion, by 
\beq
E_{\rm gap} = \frac{1}{8}E_{\rm breakdown}(J=0) = \frac{c^4 \hbar^2}{8  G_N^{1/2}|Q|^3}
\eeq
 and the large ground state degeneracy now survives the extremal limit
 \bea
 \frac{S(T)}{k_B} &\approx& \frac{\pi Q^2}{c \hbar}+ c_{\rm log} \log \left(\frac{\pi Q^2}{c \hbar}\right)+ O(1)\nonumber\\
 &&+O(e^{-E_{\rm gap}/k_B T}) \label{eqn:susyS}
 \ea
This is an example of an expected violation of the third law -- the ground state degeneracy is protected by supersymmetry anyways. The first line contains temperature independent corrections to the ground state entropy depending on the coefficient $c_{\rm log}$ which was computed by Sen and matched with string theory in multiple situations \cite{Sen:2008vm,Sen:2011ba}. The second line includes the leading temperature-dependent correction, which are exponentially suppressed thanks to the gap. Extremal black holes therefore do exist, only when supersymmetric. 

It is still an open question to elucidate the gravitational description of these supersymmetric black hole microstates. Some proposals were put forward by the fuzzball program \cite{Bena:2022rna}. Signatures of those microstates in matter correlators were proposed in \cite{Lin:2022rzw}. As another example, surprisingly, in some cases the gravitational path integral when combined with supersymmetric localization reproduces exactly the ground state entropy \cite{Banerjee:2009af,Dabholkar:2011ec,Iliesiu:2022kny} and not only the large charge limit implicit in eqn. \eqref{eqn:susyS}.

To conclude, JT gravity played a crucial role in uncovering the correct spectrum of near-extremal black holes. It has also provided a fruitful solvable model of quantum gravity that has clarified various other quantum aspects of black hole physics such as quantum chaos, the relation with random matrix models, and the evaluation of the entropy of Hawking radiation for an evaporating black hole. This is therefore a vibrant area of research and several fundamental questions still remain open.

\begin{center}
{\bf Acknowledgements}
\end{center}

\vspace{-2mm}

We thank E. Witten for discussions and M. Heydeman, L. Pando Zayas, S. Wadia and E. Witten for comments on the draft. GJT is supported by the Institute for Advanced Study and the National Science Foundation under Grant No. PHY-2207584, and by the Sivian Fund.

\bibliographystyle{utphys2}
{\small \bibliography{references}{}}

\end{document}